\documentclass[twocolumn,amsmath,amssymb,aps,pra,]{revtex4-1}
\usepackage{amsmath}
\usepackage{blkarray}
\usepackage{braket}
\usepackage{graphicx}
\usepackage{dcolumn}
\usepackage{bm}
\usepackage[style=base]{caption}
\usepackage{subcaption}
\usepackage{stackengine}
\begin{document} 

\preprint{APS/123-QED}

\title{Loschmidt echo and Momentum Distribution in a Kitaev Spin Chain}
\author{Vimalesh Kumar Vimal $^1$}
 \email{vimalkv@iitk.ac.in}
 \author{H. Wanare $^1$}
 \email{hwanare@iitk.ac.in}
 \author{V. Subrahmanyam $^2$}%
 \email{vmani@iitk.ac.in}
\affiliation{$^1$Department of Physics,  Indian Institute of Technology, Kanpur 208016, India,}
\affiliation{$^2$School of Physics, University of Hyderabad                       
Gachibowli, Hyderabad-500046, India}

\date{\today}

\begin{abstract}
We investigate the Loschmidt echo in a one-dimensional spin chain having Kitaev-type interaction in constant and kicked magnetic fields. The Loschmidt echo for the initial states having different magnon excitations shows long-time revivals for smaller chains and has short-time revival peaks for the longer chains. The system near the critical point shows peculiarly long-time revival peaks of the Loschmidt echo for relatively larger chains. The presence of a magnon in the initial state affects the Loschmidt echo revival peaks. The momentum distribution function exhibits maxima for a few momenta that are associated with the momentum of the magnon excitation present in the initial states. The probability maxima decay as $O(1/N)$ with the system size. For the Hamiltonian with kicked magnetic fields, the Loschmidt echo depends on the kick period. For a special kick period, the Loschmidt echo shows no evolution at all irrespective of the system size.
\end{abstract}

\maketitle

\section{Introduction}
The recent experimental advancement in ultracold atoms trapped on the optical latices\cite{BlochI, Belsley} has created a renewed interest to explore the dynamics of quantum systems, particularly, using quantum quenches\cite{Polkovnikov, Mitra, Zurek, Mistakidis, Mistakidis1}. For closed quantum systems, the quantum quenching leads to a unitary evolution which can be determined by the Loschmidt echo analysis\cite{Lupo, Piroli}.
For the quantum systems quenched to the critical point, the dynamics of the finite chains have periodic revival peaks structures which decrease with the increasing system size\cite{Quan, Yuan, Rossini, Happola, Montes, Igloi, Jafari, Jafari1, Najafi}. The enhanced decay of the Loschmidt echo(LE) can be considered as the witness of the quantum phase transition\cite{Haikka, Bayat, Jafari2, Bayat2}. The singularities of LE can also give the signature of strongly localized phases\cite{Leonardo}. The extensions of the Loschmidt echo have also been used to study the information scrambling\cite{Chenu,Lin}. In this process, the local information of the system disperses to the non-local degrees of the freedom throughout the system\cite{Landsman, Joshi, Blok, MiX, Braumuller, Sreeram, Shukla}. The Loschmidt echo can be computed by taking the overlap of the prior to and after the quenched state that can be tuned using the Hamiltonian parameters. The rate function defined using the LE has been studied extensively to trace the signal of dynamical phase transition in many quantum systems\cite{Heyl, Rylands, Vajna, Andraschko, Kriel, Canovi, bam, Heyl1, Lacki, Piccitto, Kyaw, Yu, Halimeh, Syed}. In recent times, the connection between the quantum quenches and the topological properties and topological edge states has also been investigated. The topological systems, especially the topological superconductor, have been investigated in detail\cite{Hasan, Qi, Alicea}, and have been shown to be quite robust to the quantum quenching\cite{Tsomokos, Hal, Nico}.

We study the Loschmidt echo in a Kitaev spin chain in one dimension\citep{Vimal1, Vimal2, Subrahmanyam} when the Hamiltonians are set at the different global parameters. We will study the behaviour of the Loschmidt echo both near and away from the critical region. We will examine initial states with no magnon excitation and initial states with one-magnon excitations.

 We also study the momentum distribution of an excited magnon for this model within the framework of the Loschmidt echo. The momentum distribution has been studied in different scenarios. The momentum distribution in spinless bosons\cite{Papenbrock} and in spin$-1$ bosons\cite{Deuretzbacher} in one-dimension have been studied previously. Also,  the signature of the Fulde-Ferrel-Larkin-Ochnikov(FFLO) phase can be seen in the momentum distribution function of the trapped one-dimensional Fermi gases\cite{Feiguin, Casula}. The momentum distribution function can give information about the spatial distribution of electrons in the quasiparticle bands\cite{Hagymasi}. In the momentum space,  it can give the probabilistic distribution of a magnon excitation in the time-evolved state of the Hamiltonian. We will see the evolution characteristic of the momentum distribution coincides with the characteristic of the Loschmidt echo dynamics. Also, the peaks of the distribution function are confined to some special values of momenta, which is expected in this case.

The Loschmidt echo (LE)  measures the degree of reversibility of the system when it evolves under a Hamiltonian for a certain amount of time and evolves back using the perturbed Hamiltonian for the same amount of time. The forward and backward evolution in time is shown pictorially in Fig.1. The Loschmidt echo is defined as the square of the modulus of overlap of the two states that evolve from the same initial state under the considered Hamiltonian and Hamiltonian with perturbation\cite{Peres}. An initial state $\ket{\psi(0)}$ evolves under the Hamiltonian $H_f$ for time $t$ and then it further evolves under the Hamiltonian $-H_b$ for the same time.  
The Loschmidt echo $L(t)$ can be written as 
\begin{equation}
\textsf{L}(t)= |\bra{\psi(0)}e^{iH_bt}e^{-iH_ft}\ket{\psi(0)}|^2.
\end{equation} 
From this,  LE can be viewed as a measure of the degree of the reversibility of the dynamics. In this paper, we investigate the Loschmidt echo for three different initial states which evolve under the Hamiltonians $H_f$ and $H_b$, which are set on and off the critical point by tuning the Hamiltonian parameters. 
The paper is organized in the following form. Section II discusses the Hamiltonian setup, its eigenstates, and the state dynamics. Section III discusses the Loschmidt echo for a zero magnon initial state. The Loschmidt echo for one magnon in the momentum space is discussed in section IV. In section V, we discuss the Loschmidt echo for an initial state with a uniform probability distribution of momenta. We consider kicked magnetic field in the Hamiltonian and discuss the LE and the momentum distribution in all three initial states in section VI. We conclude the results in section VII.

\section{Eigenstates of the Hamiltonian}
We consider a system of $N$ spins in one dimension having nearest-neighbor interactions in the presence of a transverse magnetic field. The nearest spins have Kitaev-type interaction, which is $x-x$ interaction on the odd pair of sites and $y-y$ interactions are on the even pair of sites. 
The spin chain Hamiltonian considered here is the one-dimensional simplification of the two-dimensional Kitaev honeycomb lattice model. The $z-z$ interaction in the Kitaev honeycomb Hamiltonian is replaced by the uniform magnetic field term. The spin chain Hamiltonian  is given by
\begin{figure}
\includegraphics[height=2.5cm,width=6cm]{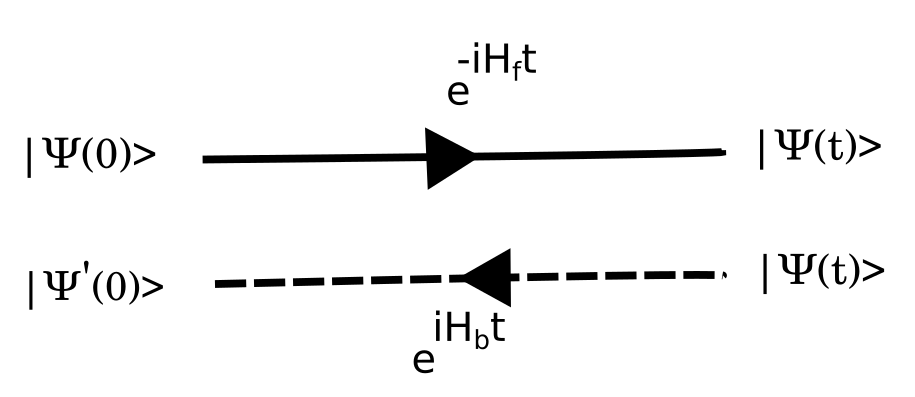}
 \caption{ Loschmidt echo}
\end{figure}
\begin{equation}
H = j_x\sum\limits_{i=odd}^{N-1}\sigma_{i}^x\sigma_{i+1}^x + j_y\sum\limits_{i=even}^N\sigma_i^y\sigma_{i+1}^y + h\sum\limits_{i=1}^N\sigma_i^z.
\end{equation}
The coefficient $j_x(j_y)$ is the strengths of the nearest neighbor interaction on the odd(even) pair of sites. The  coefficient $h$ is the strength of the uniform magnetic field in the system. The spin model can not be simplified to the Ising or the $xy$ spin chains, because a spin at any site in this model has only one, either $x$ or $y$,  direction of interaction with its next nearest neighbor. For this reason, it can not be mapped into the Ising or the $XY$ spin models. The presence of only one degree of interaction at each site adds surprising features to this model, some of which are contrary to the one-dimensional behavior. The Hamiltonian has a macroscopic degeneracy in the ground state in absence of the magnetic field. However, the ground state quantum correlation measures like the concurrence measure of the entanglement, and the quantum discord do not show an expected scaling behavior near the critical point of the system. The Hamiltonian is  diagonalisable using Jordan-Wigner fermion method, and all the eigenstates constructed and the dynamics can be studied \cite{Vimal1, Vimal2, Subrahmanyam}. We will represent the Hamiltonian parameters in the unit of $j_x$, effectively making it a two-parameter family of $r=j_y/j_x,$ and $h/j_x$.
Thus, we will represent the Hamiltonian as $H=H(r,h/j_x)$ in the following. 
We will review briefly how to study the dynamics of an initial state using this Hamiltonian. 
In the momentum space, the Hamiltonian takes the form, $H=2\sum_{q}H_q$, where the sum is over $N/4$ momentum values, $0<q<\pi/2$. Each $q$ is associated with the four momentum values, $q-\pi$, $-q,q$, and $\pi-q$. The free fermion form of $H_q$ can be written as
\begin{equation}
 H_q=\sum_{i=1}^{4}\lambda_i\xi_i^\dagger\xi_i.
\end{equation} 
Here $\lambda_i=\pm |e| \pm \sqrt{|e|^2+h^2}$ are the  eigenvalues of the mode Hamiltonian. The parameter $|e|=\frac{1}{2}\sqrt{((j_x+j_y)\cos q)^2+((j_x-j_y)\sin q)^2}$. The index $i$ in increasing order implies the increasing values of $\lambda_i$. The operators $\xi_i$ are the eigenoperators corresponding to $\lambda_i$, which can be written as

\begin{equation}
\begin{split}
\begin{bmatrix}
           \xi_{1}^{\dag} \\
           \xi_{2}^{\dag} \\
           \xi_{3}^{\dag}\\
           \xi_{4}^{\dag}
  \end{bmatrix} = 
  \begin{bmatrix}
     {\cal C}~ \quad & i {\cal S}~   & h_1 {\cal C}~ & \quad  i h_1 {\cal S}~ \\
     {\cal S}~  \quad & -i {\cal C}~ &   h_2 {\cal S}~ &\quad -i h_2 {\cal C}~ \\
     {\cal C}~ \quad &  i{\cal S}~   & h_3 {\cal C}~ & \quad  i h_3 {\cal S}~ \\
     {\cal S}~   \quad & - i {\cal C}~ &   h_4 {\cal S}~ &\quad -i h_4 {\cal C}~ \\
  \end{bmatrix}
  \begin{bmatrix}
           F_+^{\dag} \\
           G_+^{\dag} \\
           F_-^{\dag}\\
           G_-^{\dag}
  \end{bmatrix} .
  \end{split}
\end{equation} 
Here ${\cal C}=\cos\theta_q/2,\quad {\cal S}=\sin\theta_q/2$ with $\theta_q = \sin^{-1} {\frac{(1-r)\sin q}{\sqrt{((1+r)\cos q)^2+((1-r)\sin q)^2}}}$, and $h_i=h/\lambda_i$. The fermion operators $F_{\pm}={(c_{q-\pi}-c_{-q}^{\dag})}/\sqrt{2}$, and $G_{\pm}={(c_{q}-c_{\pi-q}^{\dag})}/\sqrt{2}$. Using Eq.3, all the eigenstates of the Hamiltonian can be created from the vacuum state defined as $\xi_i\ket{vac}=0$. The unnormalized ground state can be written as 
\begin{equation}
\begin{split}
\ket{g}=\prod_{0\leq q \leq \pi/2}^{}{\large [} (1-h_1)(1-h_2)+ \lbrace(1-h_1h_2)+ \\ (h_2-h_1)\cos\theta_q\rbrace c_q^{\dag}c_{\pi-q}^{\dag}-i(h_1-h_2)\sin\theta_q(c_{-q}^{\dag}c_{q}^{\dag}+\\ c_{q-\pi}^{\dag}c_{\pi-q}^{\dag})+\lbrace(1-h_1h_2)-(h_2-h_1)\cos\theta_q\rbrace c_{q-\pi}^{\dag}c_{q}^{\dag}+\\(1+h_1)(1+h_2)c_{q-\pi}^{\dag}c_{-q}^{\dag}c_q^{\dag}c_{\pi-q}^{\dag}{\large ]\ket{0000}}.
\end{split}
\end{equation}
Similarly, we can construct the excited states using the mode operators $\xi^\dag_i$ on the vacuum state for different $q$ values.
 In the absence of magnetic field there are mode operators with zero energy. The presence of the zero energy eigenoperators causes the Hamiltonian to have a  macroscopic degeneracy in its ground state. The quantum correlations also have surprising features in this model. Contrary to the Ising and $xy$ spin chains, the quantum correlations like the concurrence measure and the quantum discord do not exhibit a signal of the quantum critical point in the system. However, these correlations show maxima  at the quantum critical point\cite{Vimal1}. The dynamics of magnetisation shows a counter intuitive revivals with respect to the concurrence and the quantum discords\cite{Vimal2}. In the following sections, we will consider different initial states and investigate the Loschmidt echo and the evolution of the momentum distribution.

\section{No-Magnon Initial State }

We consider an initial state of spins completely polarized in the $-ve$ z-axis, which translates to zero-fermion state in the momentum space. The state is  written as $\ket{\psi(0)}=\ket{00..00}$. This state can be thought of as an eigenstate of the Hamiltonian in Eq.2  when the magnetic field is very large. According to the protocol discussed in Eq.1, we can write the forward evolution as
\begin{equation}
\ket{\psi(t)}=e^{-iH_ft}\ket{00..00}.
\end{equation}
\noindent
In the above expression, $H_f= H(r, h_f/j_x)$ represents the forward Hamiltonian with $h_f$, the magnetic field. It can be written as the sum of the mode Hamiltonians $H_q$(shown in Eq.3) that commute with each other, $H_f=2\sum_qH_q$.
\begin{figure}
\includegraphics[height=6cm,width=8cm]{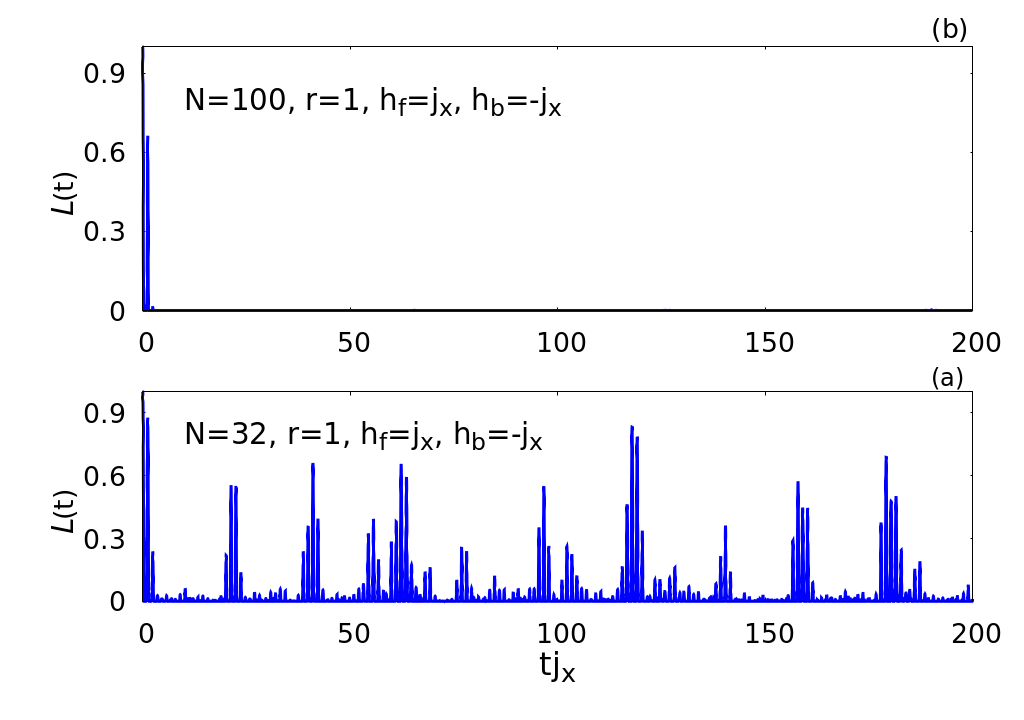}
 \caption{ Loschmidt echo as a function of time for (a) a shorter chain of $N=32$, (b) a larger chain of $N=100$. The forward evolutions in both cases are at $r=1$ and $h_f/j_x=1$, while for the reverse evolution, the magnetic field is flipped. The Loschmidt echo shows long-time revival peaks for smaller chains. For the larger spin chains, it decays exponentially and has only short-time revival peaks.}
\end{figure}
 Thus, the forward evolution can be further written as

\begin{equation}
\ket{\psi(t)}= e^{-2i\sum_qH_qt}\ket{00..00} =\Pi_{q}\ket{\phi_q(t)}.            
\end{equation} 
Here, $\ket{\phi_q(t)}\equiv e^{-2iH_qt}\ket{0000}_q$ evolves under the mode Hamiltonian $H_q$ in the forward direction. Similarly, the backward evolution happens under the Hamiltonian $H_b= H(r, h_b/j_x)$ where $h_b$ is the magnetic field for this evolution. The Hamiltonian can be written as $H_b=2\sum_qH_q'$. Here $H_q'$ are the mode Hamiltonians of $H_b$. The time-evolved state can be written as

\begin{equation}
\ket{\psi'(t)}= e^{-2i\sum_qH_q't}\ket{00..00}\equiv\Pi_{q}\ket{\phi'_q(t)},
\end{equation} 
where the state $\ket{\phi_q'(t)}=e^{-2iH_q't}\ket{0000}_q$ evolves under the mode Hamiltonian $H_q'$ in the reverse direction. Therefore, the Loschmidt echo in Eq.1 can be simplified as the square of amplitude of the overlap of the two states $\ket{\psi(t)}$ and $\ket{\psi'(t)}$, given as

\begin{equation}
\textsf{L}(t)=|A|^2,\quad A=\Pi_{q}A_q,\quad A_q=\braket{\phi'_q(t)|\phi_q(t)}.
\end{equation}
Using Eq.5 and its equivalent for the excited states, from above we can calculate the Loschmidt echo as a function of time for different magnetic field values for the forward and the reverse directions. In Fig.2, LE has been plotted when forward evolution happens under the Hamiltonian  $H_f(r=1, h_f/j_x=1)$, and the reverse evolution happens by flipping the magnetic field direction $H_b(r=1, h_b/j_x=-1)$, the local interaction $j_x$ has been set unity throughout the analysis. The Loshdmidt echo for small chains has a periodic structure in the evolution while for longer spin chains its revival peaks reduce. Fig.2(a) shows behavior of the Loschmidt echo for the smaller spin chains while Fig.2(b) shows the same for the larger chain lengths. The Loschmidt echo at $t=0$ is unity as the system is in its initial state. As time progress, the LE decays exponentially. However, it revives very quickly and a periodic structure of dynamics appears over time. In a long time evolution, the revival peaks appear only for the smaller chain lengths, which can be seen in Fig.2(a). For up to $N=44$, the LE shows a revival peaks of the half of its maxima in the long time evolution but as the length of the chain increases, these peaks start vanishing. In Fig.2(b), for $N=100$, only one peak is significant which appear soon after the evolution and no further significant revival peaks are seen in the long time evolution. For the larger spin chain, the $N/4$ possible $A_q$ functions in Eq.9 go out of phase soon after the evolution thus making the Loschmidt echo difficult to revive in long-time evolution. We can also analyze the LE behavior at different $r$ values. At $r=0$, $h_f=1$, $h_b=-1$, the LE shows a periodic behavior as a function of time. This may be for a reason that at $r=0$, all the $A_q$  functions are periodic and remain in phase over time. As we increase $r$ from 0 to 1, the LE loses its periodic nature and falls to zero, and remains grounded.
\begin{figure*}[t]
 {(a) \hskip 6.5 cm (b)}\\
\begin{subfigure}{1.0\textwidth}
\includegraphics[width=0.3\linewidth, height=5cm]{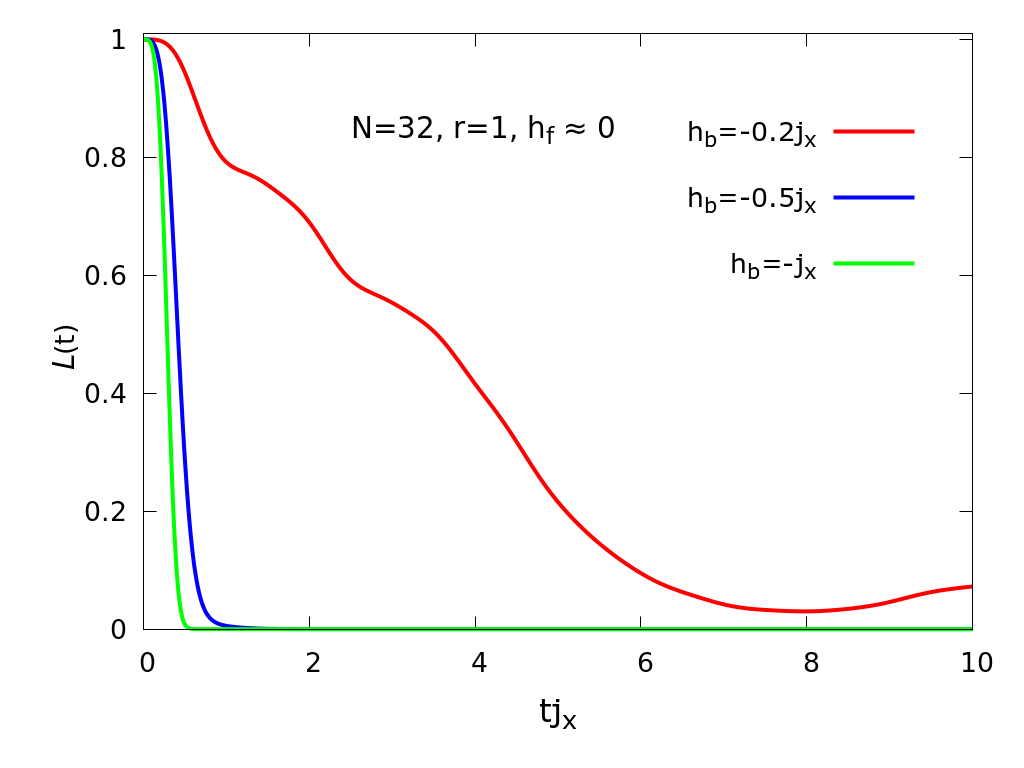} 
\includegraphics[width=0.3\linewidth, height=5cm]{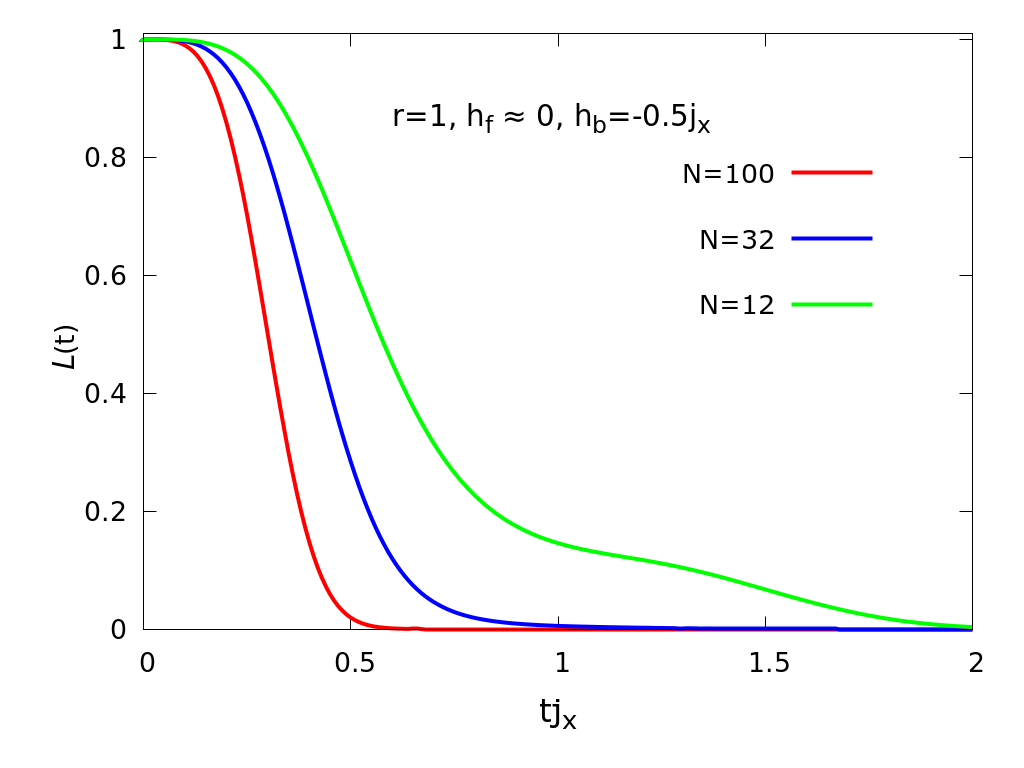}
\includegraphics[width=0.3\linewidth, height=5cm]{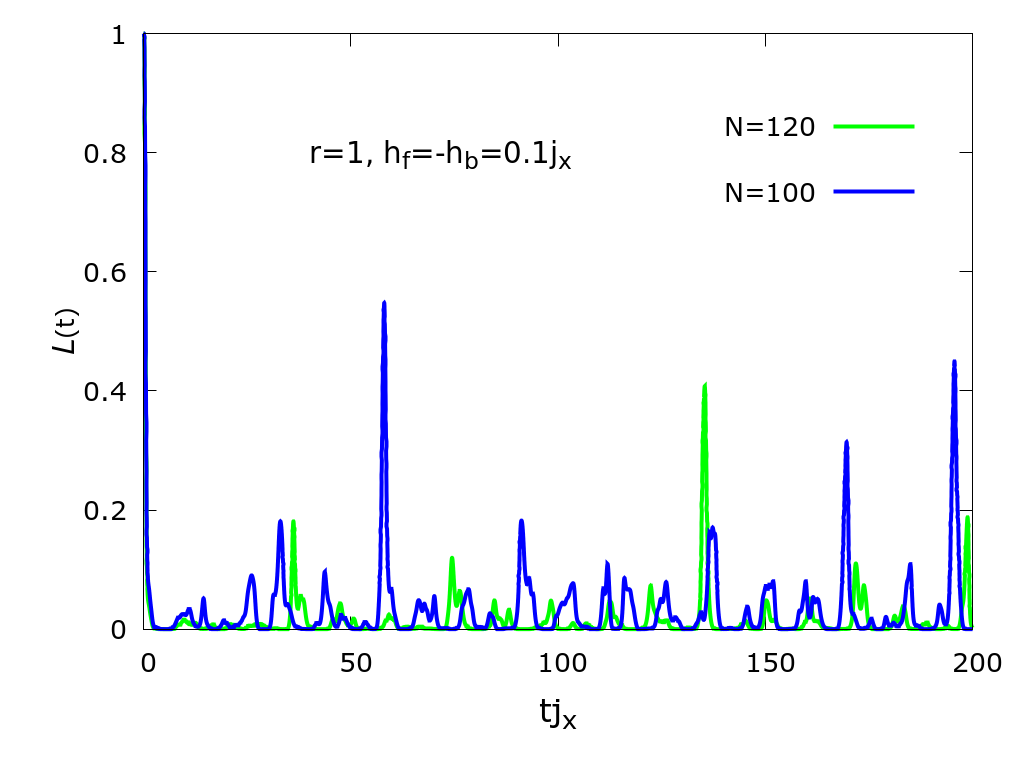}
\end{subfigure}
 \caption{Loschmidt echo near the critical point. (a) For a small chain of $N=32$ at $r=1$, $h_f=0$, and different values of $h_b$. The Loschmidt echo decays very sharply in the higher magnetic field in the reverse evolution. (b) At $r=1$, $h_f=0$, and $h_b=-0.5j_x$ for different spin chains. The LE falls more quickly and does not show revival for the longer spin chains. The revival peaks are present for the smaller chains which we do not show in the plot. As the length of the spin chain increases beyond $N=16$, the revival is not possible. (c) For  $h_f=-h_b =0.1j_x$, the Loschmidt echo shows peaks after a long-time evolution even for larger spin chains, $N=100$ and $N=120$. Beyond this length, the revivals peaks fall quickly. For the smaller chains, the revivals peaks are periodic with higher amplitudes. 
 }

\end{figure*}
When the evolution in either direction happens under the Hamiltonian near the critical point the revivals peaks of Loschmidt echo are gone completely for even smaller spin chains. Fig.3 shows the behavior of LE for the different spin chains in such a scenario. In Fig.3(a) the LE is plotted for a spin chain of $N=32$. We set $r=1$ and the forward magnetic field,$h_f=0$, and consider different magnetic fields $h_b$ when system evolves backward in time. We can see as we increase $h_b$, the Loschmidt echo falls very sharply and never revives. In Fig.3(b), the LE has been plotted at fixed $h_b=-0.5j_x$ keeping rest of the parameters same as they are in Fig.2(a). We can see the LE shows polynomial decay for small spin chains while it shows a sudden fall for large spin chains. Therefore we show only short-time dynamics for these two cases. In Fig.3(c), we have shown the behviour of LE for larger lengths $N=100$ and $N=120$ in the critical region $h_f=0.1 j_x$ and $h_b=-0.1j_x$. In this case, the Loschmidt echo exhibits revival peaks after long-time evolution. However, these peaks sharply fall beyond these lengths. For smaller size chains, the LE exhibits periodic revival peak with higher amplitudes. The quick revival peak that we see in a non-critical regime is not present in this case. However, the long-time revivals of the Loschmidt echo is pronounced only at $h_f=-h_b =0.1j_x$, which disappear as we tune the magnetic field even slightly to $h_f=-h_b =0.15j_x$. The magnetic fields near the critical point will give a very fluctuating behavior as expected. Also, in this case, the revival does not occur at near the same time for the different lengths of the chain as it occurs for the short-time revival peak in the non-critical Hamiltonian cases. This may induce the behavior that the values of the revival peaks as a function of $N$ do not show any certain characteristics in this case. However, for chains larger than $N=100$, the peaks fall very quickly and become insignificant which is the general character of the system.
In a different scenario of a two-level system surrounded with the Ising type spin chain, the time of the revival peaks has been shown proportional to the length of the chain\cite{Quan}. They also show that the enhanced decay of LE can be used to witness quantum criticality.

\section{One-Magnon Initial State With a Definite Momentum }
In the last section, we have studied the even-number magnon state starting from the zero-magnon initial state. In this section, we will consider the odd-number magnon state in evolution, starting with a one-magnon initial state. Let us consider an initial state that has only one magnon with a definite momentum $q$,  given by
 
\begin{equation}
\ket{\psi_q(0)}=c_q^{\dag}\ket{00..00}.
\end{equation}
This state will evolve into a superposition of  states with different odd number of magnons.
The evolution of this state can be written using the evolution of $c_q^{\dag}$ under the Hamiltonian $H_f$ defined for Eq.6 as

\begin{equation}
\ket{\psi_q(t)}= c_q^{\dag}(-t) e^{-iH_ft}\ket{00..00},\\[2pt]         
\end{equation}
where the time evolution of the creation operator is written as

\begin{equation}
c_q^{\dag}(-t)=e^{iH_ft}c_q^{\dag}e^{-iH_ft}.
\end{equation}
The time evolution term $e^{-iH_ft}\ket{00..00}$ in Eq.11 can be computed using Eq.7. The state $\ket{\psi_q(t)}$ can further evolve under the Hamiltonian $H_b$. The time-evolved creation operator $c_q^{\dagger}(-t)$ of mode $q$ is a function of the momentum values $\lbrace q-\pi,-q,q,\pi-q \rbrace $. There are $N/4$ such momentum values allowed for the Hamiltonian, which each having has four modes. 
Therefore, $C(k,q)$ can have three more possibilities with $\lbrace k=q-\pi, -q, \pi-q\rbrace$. 
Thus, $c_q^{\dagger}(-t)$ can be written in terms of the other associated momenta operators as

\begin{equation}
c_q^{\dagger}(-t)=\beta_1 c_{q-\pi}^{\dag}+\beta_2 c_{-q}+\beta_3 c_q^{\dag}+\beta_4 c_{\pi-q},
\end{equation}
where $\beta_j=\sum_{i=1}^{4}e^{-2i\lambda_i t}\Gamma_{i3}^{*}\Gamma_{ij}$. $\lambda_i$ are the eigenvalues of the Hamiltonian $H_q$ and the $\Gamma$ matrix is written as

\begin{equation}
\begin{split}
\Gamma =
  \begin{bmatrix}
     (1+h_1){\cal C} & (1-h_1) {\cal S}~ & i(1+h_1) {\cal C}~ &   i(1-h_1) {\cal S}~ \\
     (1+h_2){\cal S} & (1-h_2) {\cal C}~ & -i(1+h_2) {\cal S}~ & -i(1-h_2) {\cal C}~ \\
     (1+h_3){\cal C} & (1-h_3) {\cal S}~ & i(1+h_3) {\cal C}~ & i(1-h_3) {\cal S}~ \\
     (1+h_4){\cal S} & (1-h_4) {\cal C}~ & -i(1+h_4) {\cal S}~ &-i(1-h_4) {\cal C}~ \\
  \end{bmatrix}.
  \end{split}
\end{equation} 
Thus the time-evolved mode operators in Eq.12 are functions of all the four momenta operators associated with the corresponding mode. During the course of the evolution, the probability distribution of the momentum may change. Therefore, we can overlap the final evolved state with the same or a different momentum state defined in Eq.10. This can be written by defining a probability distribution function $P_q(k,t)$, which is essentially a momentum distribution function of the momentum $k$ in the time-evolved state which has the initial state with a definite momentum $q$. This can be rewritten as

\begin{equation}
P_q(k,t)= \bra{\psi^{\prime}(t)}c_k'(-t)c_q^{\dag}(-t) \ket{\psi(t)}.
\end{equation}
where the $\ket{\psi^{\prime}(t)}$ is defined in Eq.8. 
The operator $c_k'(-t)= e^{iH_bt}c_q^{\dag}e^{-iH_bt}$ defined for the backward evolution gives the time evolution of the momentum $k$ under the Hamiltonian $H_b$.
The time evolved momentum operator $c_q^\dag(-t)$ acts only on $\ket{\phi_q(t})$, leaving other mode states of $\ket{\psi(t)}$ unaffected. Similarly, $c_k'(-t)$ acts only on $\bra{\phi_q'(t})$ leaving other modes state of $\bra{\psi^{\prime}(t)}$ unaffected. Therefore, we can rewrite the probability distribution function of a momentum $k$ as 

\begin{equation}
P_q(k,t)= |A/A_q|^2|C(k,q)|^2,
\end{equation}
where $A$ and $A_q$ are defined in Eq.9 and the expression

\begin{equation}
C(k,q)=\bra{\phi_q^{\prime}(t)}c_k'(-t)c_q^{\dag}(-t) \ket{\phi_q(t)}.
\end{equation}
\begin{figure}
\includegraphics[height=5cm, width=7cm]{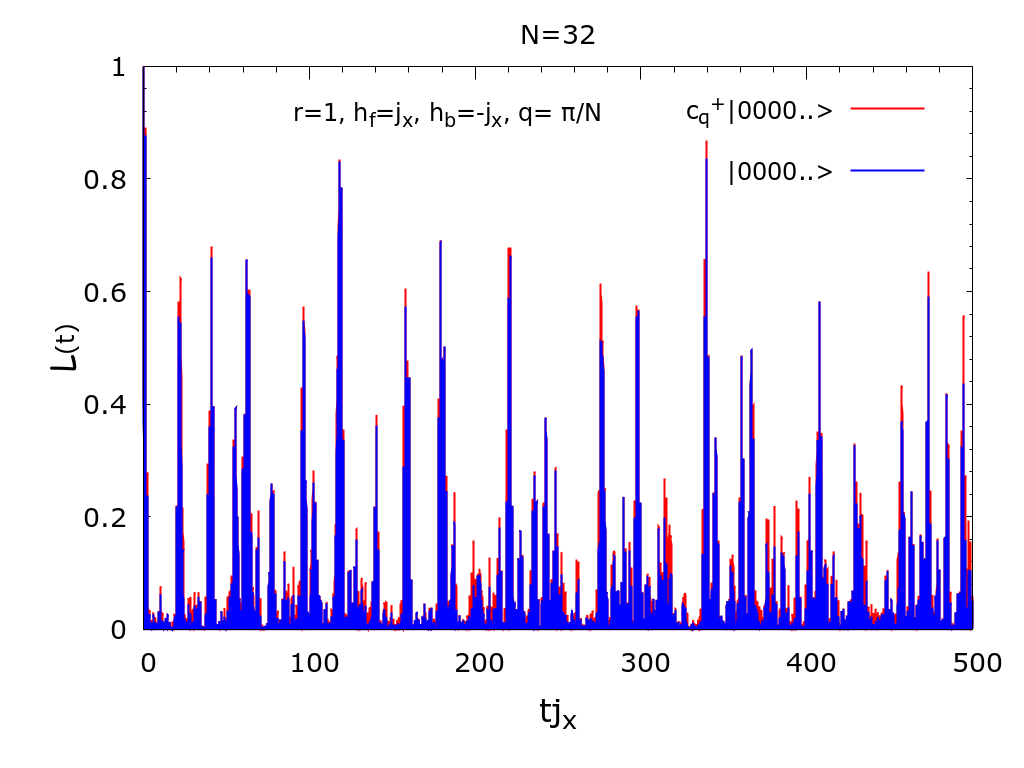}
\caption{Loschmidt echo for the one magnon initial state with a definite momentum. The plot is for $N=32$ at $r=1$, $h_f=j_x$, and $h_b=-j_x$. The momentum present in the initial state increases the revival peaks by a smaller amount. However, the increase is distinguishable only for smaller chains. }
\end{figure}
\noindent The probability distribution function  for $k=1$  is just the Loschmidt echo. Therefore, the Loschmidt echo expression is given by
\begin{equation}
\textsf{L}(t)= |A/A_q|^2|C(q,q)|^2,
\end{equation}  
We can see in the above that the presence of a magnon with a fixed momentum value in the initial state affects only the mode state associated with that momentum. Therefore, the Loschmidt echo for the larger spin chains does not show the effect of the excitation in the initial state and has a similar result to the Loschmidt echo for the no magnon initial state. However, it has a significant impact on the smaller chains. In Fig.4, we plot the Loschmidt echo for a spin chain of $N=32$ for the short and the long time evolution. The Hamiltonian parameters are set as $r=1, h_f=j_x, h_b=-j_x$. The revival peaks get a little stronger by the presence of magnon in the initial state. However, when the Hamiltonian is set in the critical zone for either direction of the evolution the revival characteristic is lost and does not show similar behavior of the Loschmidt echo having no-magnon excitation in the initial state. To see the impact of the presence of one magnon excitation in the initial state on the Loschmidt echo, we plot the time-averaged LE as a function of spin length in Fig.5. The average value of Loschmidt Echos is calculated for a long-time evolved function up to $tj_x=500$. This is necessary for relatively smaller chains. However, for larger chains, $tj_x=5$ is a sufficient time of evolution to calculate the average LE as it falls sharply in a small time and remains zero in further evolution. The plot shows the vanishing gap between LE in two cases as the spin system gets bigger. The stronger revival peaks for the one magnon initial state places the time-averaged $L(t)$ on the top in the plot. We can see for $N=40$ and beyond, the two LE values merge completely, which shows the diminishing effect of a one magnon excitation in the initial state.
\begin{figure}
\includegraphics[height=5cm, width=7cm]{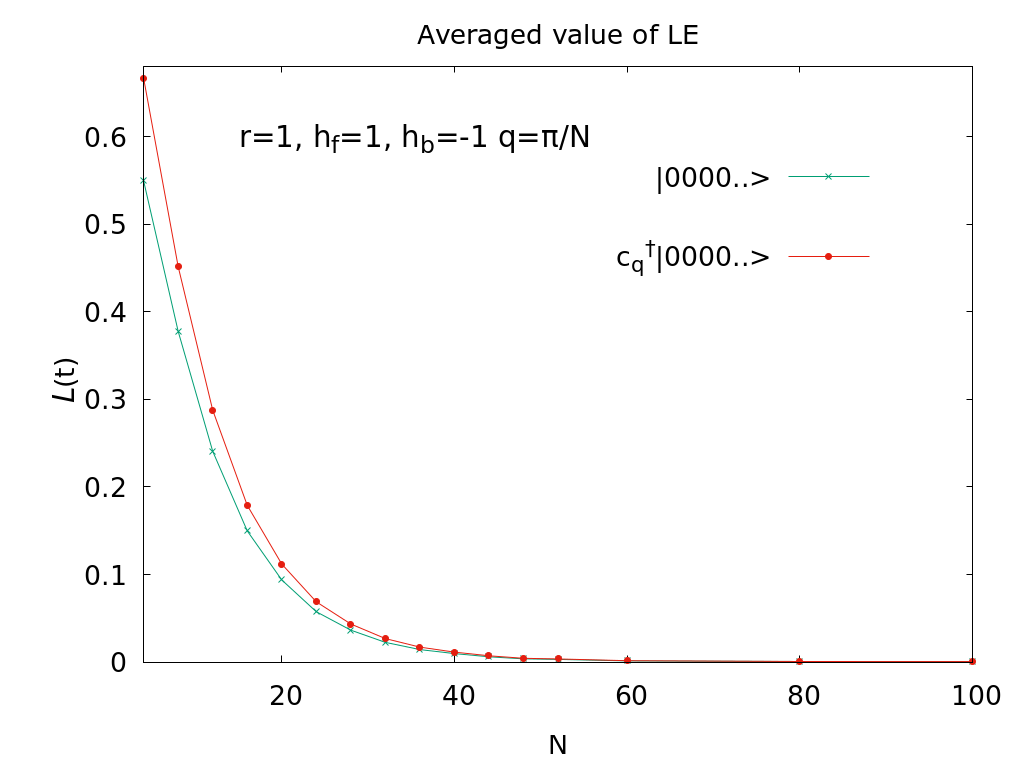}
\caption{Averaged values of the Loschmidt echo as a function of chain lengths for two different initial states. The system parameters are $r=1$, $h_f=j_x$, and $h_b=-j_x$. One magnon initial state with a definite momentum has frequent revival peaks which increases the average value of the Loschmidt echo as compared to the Loschmidt echo with no magnon initial state. }
\end{figure}

We have seen that for the larger chain lengths, the one magnon initial state with a definite momentum does not have an impact on the characteristic of $L(t)$. Therefore, we consider $N=32$ to see the effect produced by it. In Fig.6 we plot the probability distribution function for the two different modes values separately. In Fig.6(a), we take $q=\pi/N$ for $N=32$. We show the results of four values of $k=q-\pi,-q,q, \pi-q$. The probability distribution function $P_q(k,t)$ gives a non zero distribution only for $k=q$ and it always gives zero for $k\neq q$ even if it belongs to the same mode, i.e., $k=q-\pi,-q,\pi-q$. The probability distribution function remains the same if we shift $q\rightarrow \lbrace q-\pi,-q,\pi-q\rbrace$ and take $k=q$(this is not shown in the plots). This may also be the reason why the probability distribution function goes to zero when $k\neq q$ even within the same mode. The zero probability distribution function means the time-reversal mode state $\ket{\phi_q'(t)}$ and the time evolved state $\ket{\phi_q(t)}$ remain orthogonal through the evolution. In Fig.6(b) we choose the last mode value given by $q=(N-2)\pi/2N$ to show the $q$ dependence of the Loschmidt echo on momentum. As compared to Fig.6(a), we can see the magnitude of the revival peaks depends on the momentum values chosen, However, the peaks of the Loschmidt echo appear at the same time for the different momentum.  Also, the Loshchmidt echo doesn't change if we change the sign of momentum in the initial state. This is because $q$ and $-q$ fall in the same mode of the Hamiltonian and we know that the probability distribution functions $P_q(q,t)$ and $P_{q-\pi}(q-\pi,t)$ have the same characteristic.  
\begin{figure}
\center
\includegraphics[height=7cm, width=8cm]{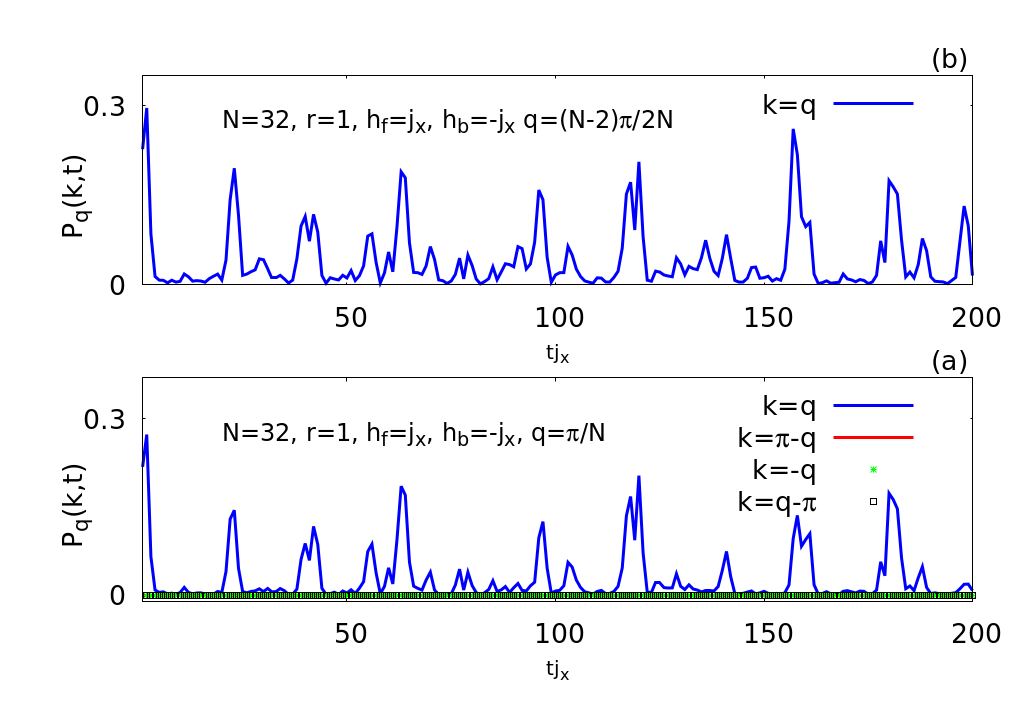}
\caption{Window averaged plots of the probability distribution functions as a function of time for different initial momentum values (a) $q=\pi/N$ and (b) $15\pi/N$, for a spin chain of $N=32$ at $r=1$, $h_f=j_x$, $h_b=-j_x$. Window length is equivalent to the 100 time values at $\delta t=0.01j_x$ apart. The probability distribution for $q\neq k$ gives zero, while for $q=k$ cases, which is equal to the Loschmidt echo, it shows the periodic revivals. For different $q$ in the initial states, the the revival peaks of the probability distributions are of different magnitudes but they occur at the same time of the evolution as shown in (a) and (b).}
\end{figure}

\section{One Magnon Initial State with Uniform probability Distribution}
The momentum distribution function of the one magnon initial state shows a distribution only for the same momentum present in the initial state. This opens the question that what would be the probability distribution of momentum if the initial state has the excitation of more than one momentum. To investigate this, we consider the initial state to be a one magnon state with the magnon localised in real space, i.e., $\ket{\psi_1(0)}=c_1^{\dag}\ket{00..00}$. In the momentum space, this is an equally probable state for all the allowed momenta of the system. In this section, we consider such an initial state and let this state evolve under the Hamiltonian $H_f$. The time-reversal state is obtained under the Hamiltonian $H_b$. The Loschmidt echo is the square of the overlap of two wave functions. For the probability distribution function analysis, we take overlap of the forward evolved state with a fixed momentum $c_k$ in the initial state and compute the probability distribution of momentum $k$ which may or may not be equal to $q$. Thus, the initial state is written as
\begin{figure}
\includegraphics[height=6cm, width=7cm]{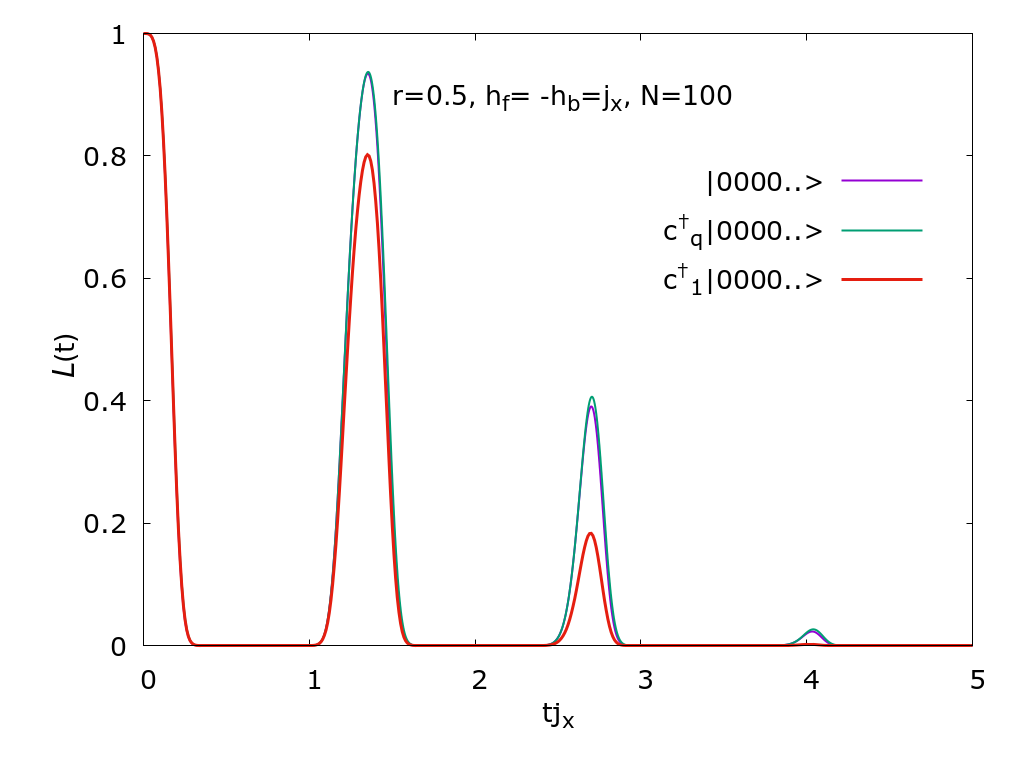}
\caption{Loschmidt echo in all the three cases of initial states for $N=100$ at $r=0.5$, $h_f=j_x$, $h_b=-j_x$. For the two initial states $\ket{\psi(0)}$ and $\ket{\psi_q(0)}$, the Loschmidt echo almost overlaps on each other. For the initial state of equally probable momenta, it shows revival peaks of smaller magnitude as compared to the rest two cases.}
\end{figure}

\begin{equation}
\ket{\psi_1(0)}= \frac{1}{\sqrt{N}}\sum_q e^{-iq}c_q^{\dag}\ket{00..00}
\end{equation}  
The evolution of the state can be given by 
\begin{equation}
\ket{\psi_1(t)}= \frac{1}{\sqrt{N}}\sum_q e^{-iq}c_q^{\dag}(-t)\ket{\psi(t)},
\end{equation} 
where $\ket{\psi(t)}$ is the state written in Eq.7. The backward evolution of this state under the Hamiltonian $H_b$ for the same amount of time and taking the overlap with the state $c_k^{\dag}\ket{00..00}$ defines the probability distribution function $P(k,t)$ of the momentum, $k$. We also call this function as momentum distribution function. This is written as

\begin{equation}
P(k,t)= \frac{1}{N}|\sum_qe^{-iq}\bra{\psi^{\prime}(t)}c_k'(-t)c_q^{\dag}(-t) \ket{\psi(t)}|^2.
\end{equation}
This can further be simplified by defining $B(k,q)=\bra{\psi^{\prime}(t)}c_k'(-t)c_q^{\dag}(-t) \ket{\psi(t)}$. We can rewrite this function as

\begin{equation}
B(k,q)=|A/A_q|C(k,q).
\end{equation}
And the probability distribution function can now be written as

\begin{equation}
\begin{split}
P(k,t)= \frac{1}{N}|A/A_k|^2|C(k,k)-C(k,k-\pi)+\\C(k,-k)e^{2ik}-C(k,\pi-k)e^{2ik}|^2.
\end{split}
\end{equation}
\begin{figure}
\includegraphics[height=6cm, width=8cm]{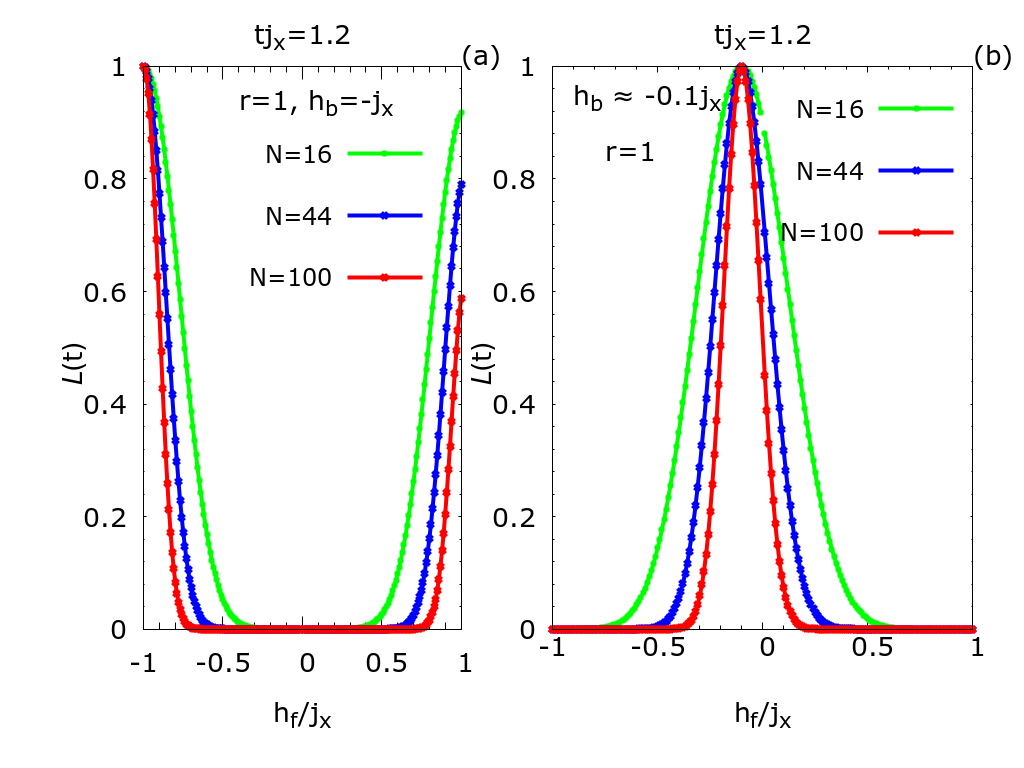}
\caption{LE as a function of magnetic field at a fixed time $tj_x=1.2$ for the different sizes of the system. (a) We consider $r=1$ and $h_b=-j_x$. The rise of the LE near $h_f=j_x$ shows that the system far away from criticality has the revival peak, which decreases with the system size. (b) We take $h_b=-0.1j_x$. When the magnetic field in the backward evolution approaches towards criticality, the revival of the Loschmidt echo is not possible.}
\end{figure}
Also, the Loschmidt echo can be written as

\begin{equation}
\textsf{L}(t)= \frac{1}{N}|\sum_{q,k}e^{-iq}e^{ik}\bra{\psi^{\prime}(t)}c_k'(-t)c_q^{\dag}(-t) \ket{\psi(t)}|^2
\end{equation}
This can further be simplified in a similar fashion as the probability distribution function. Using the momentum values selection for $k$ and $q$.

\begin{equation}
\begin{split}
\textsf{L}(t)= \frac{1}{N^2}|\sum_k[B(k,k)-B(k,k-\pi)+\\B(k,-k)e^{2ik}-B(k,\pi-k)e^{2ik}]|^2
\end{split}
\end{equation}
To simplify it further, we make use of Eq.22 and write it using momentum values, $k-\pi$, $-k$, $k$, and $\pi-k$ as

\begin{equation}
\begin{split}
\textsf{L}(t)= \frac{1}{N^2}|A|^2|\sum_{0\leq k\leq \pi/2}\frac{1}{A_k}[C(k,k)+C(k-\pi,k)+\\C(-k,k)+C(\pi-k,k)-C(k-\pi,k-\pi)-\\C(-k,k-\pi)-C(k,k-\pi)-C(\pi-k,k-\pi)+\\e^{2ik}(C(k-\pi,-k)+C(-k,-k)+\\C(k,-k)+C(\pi-k,-k)-C(k-\pi,\pi-k)-\\C(-k,\pi-k)-C(k,\pi-k)-\\C(\pi-k,\pi-k))]|^2
\end{split}
\end{equation}
\begin{figure*}
\includegraphics[height=6cm, width=7cm]{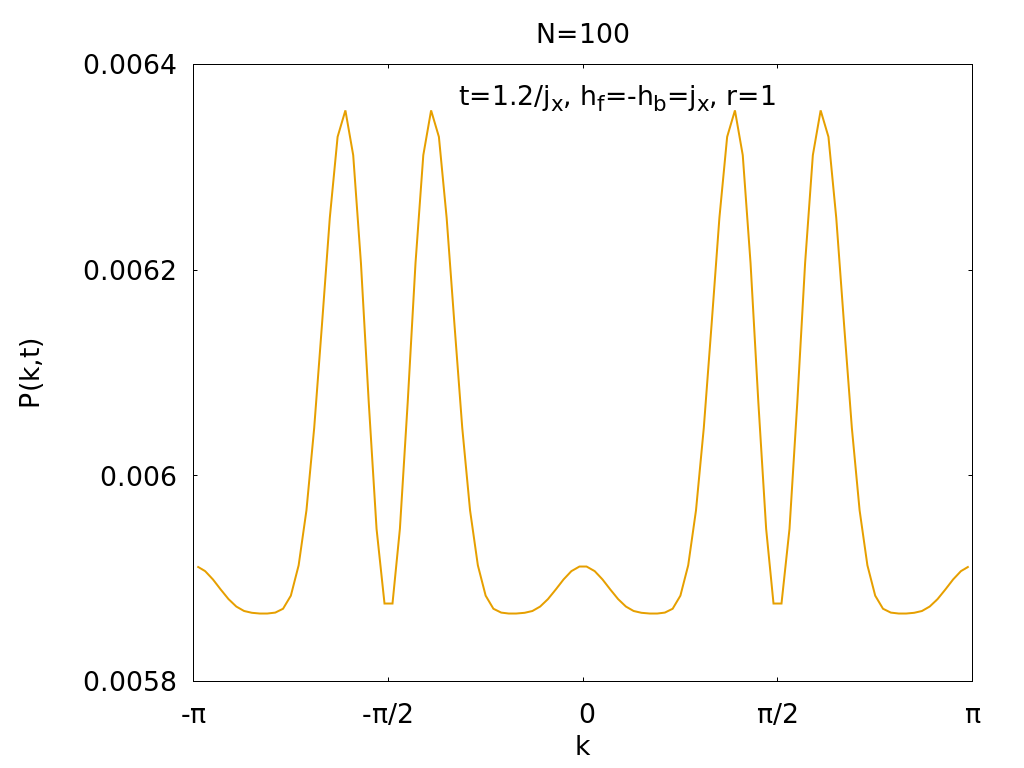}
\includegraphics[height=5.8cm, width=7cm]{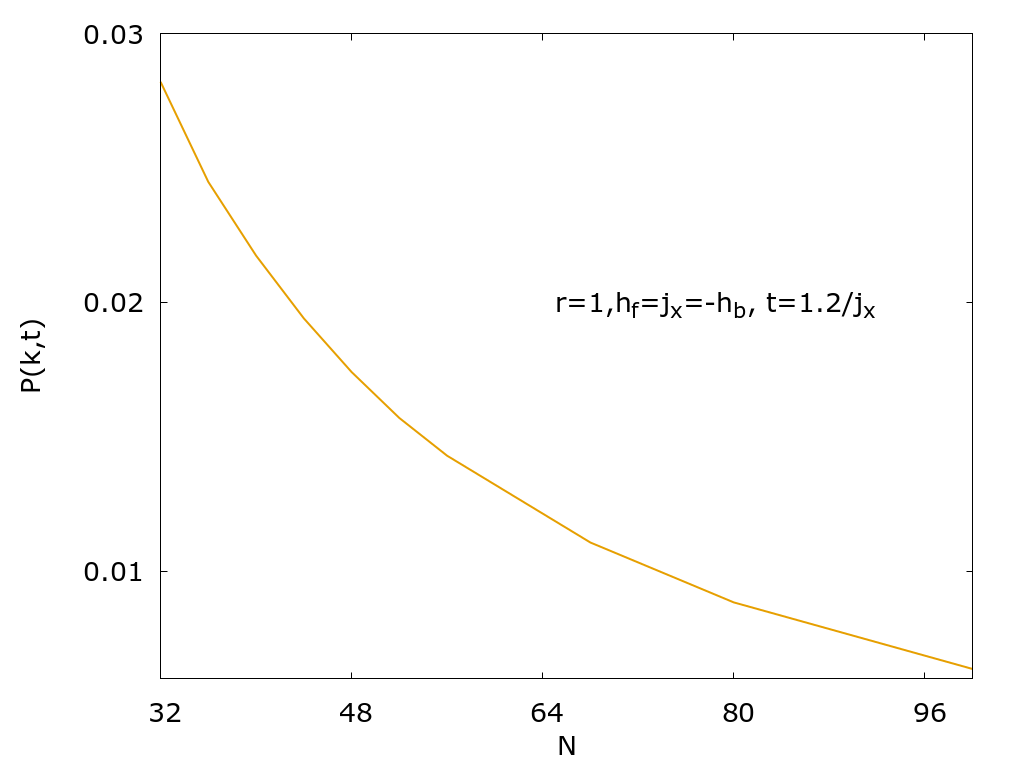}
\caption{(a) The probability distribution function as a funtion of $k$ in full range at a specific time $j_xt=1.2$ for $N=100$. The Hamiltonian parameters are $r=1$, and $h_f=1=-h_b$. the initial state for the state evolution is $\ket{\psi(0)}=c_1^{\dag}\ket{00..00}$. The probability distribution function depends on the value of $k$. It approaches to maxima for four values of $k$. These four peaks structure basically shows that the probability distribution functions of $P(k-\pi,t)$, $P(-k,t)$, $P(k,t)$, and  $P(\pi-k,t)$ as a function of time are the same for a spin chain. (b) The peaks of probability distribution as a function of $N$. The peak falls as $1/N$ with size of the chain.}
\end{figure*}
Using Eq.23 and Eq.26, we calculate the momentum distribution function and the Loschmidt echo respectively, for the initial state with a flat momentum distribution. For this state, the characteristic of the Loschmidt echo does not change as compared to the no magnon initial state or the one magnon initial state with a definite momentum. In Fig.7, we plot the results for $N=100$ spins. In this case, the Loschmidt echo does not revive as much as it revives in the rest of the two cases. For no-magnon and one-magnon with definite momentum initial states, the Loschmidt echo plots almost overlap with each each other. However, the presence of all momenta in the initial state reduces the peaks strengths. For the smaller chains, the difference in the behavior of the Loschmidt echos is a little more pronounced in the long-time evolution. For larger spin chains, the Loschmidt echo remains zero after a certain time. The Loschmidt echo can become zero if the time evolved states in the forward and reverse evolution become orthonormal. This can alternatively be achieved by tuning the magnetic field and letting the state evolve to some time. We show the behavior of $L(t)$ as a function of the magnetic field in Fig.8 for the different lengths of the spin chain. In this result, we change the magnetic field in one direction of the evolution keeping the rest of the parameters fixed. At a specific time $tj_x=1.2$, at which the revival peak appears, we show the Loschmidt echo as a function of the magnetic field in the forward direction with a fixed value of it in the reverse direction. In Fig.8(a) we plot at $h_b=-j_x$. We can see the Loschmidt echo falls to zero when $h_b\rightarrow 0$ but revives when $h_b\rightarrow j_x$. The revival strengths depend on the length of the chain. In Fig.8(b), we plot at $h_b=-0.1j_x$. In this case, we do not see the revival of the Loschmidt echo at all. This justifies the results that we have plotted for the Loschmidt echo so far. It implies that when the system is comfortably far away from the criticality in either direction of the evolution, it can have revival peaks even at the very large system size. However, when the system is set near criticality in one of the directions of the evolution, the revival of the Loschmidt echo is not possible even at lower lengths of the spin chain. 

We have shown the momentum distribution function of Eq.23 in the full range of $k$ in Fig.9(a). The four peaks have the same value and they correspond to the momentum values $q-\pi$, $-q$, $q$, and $\pi-q$. In the initial state with equally probable momenta, overlapping with a state with a definite momentum can have contributions from the four modes states of the state $\ket{\psi_1(t)}$ in Eq.20, which can have the same momentum. For this reason, we have four peaks in the momentum distribution function plotted against $k$. The behavior of the momentum distribution depends on the specific time of evolution as well as the length of the spin chain. We have considered the time $tj_x=1.2$, where the Loschmidt echo shows its first revival peak. The height of the peaks in the probability distribution decreases as the length of the spin chain increases. In Fig.9(b) the fall of the peaks is plotted as a function of $N$. The height of the peaks follow a power law as $P(k,t) = 2.177231*N^{-1.251044}$. We do not show $P(k,t)$  as a function of time as it shows a similar pattern to the Loschmidt echo.
\begin{figure*}
\includegraphics[height=10cm, width=14.5cm]{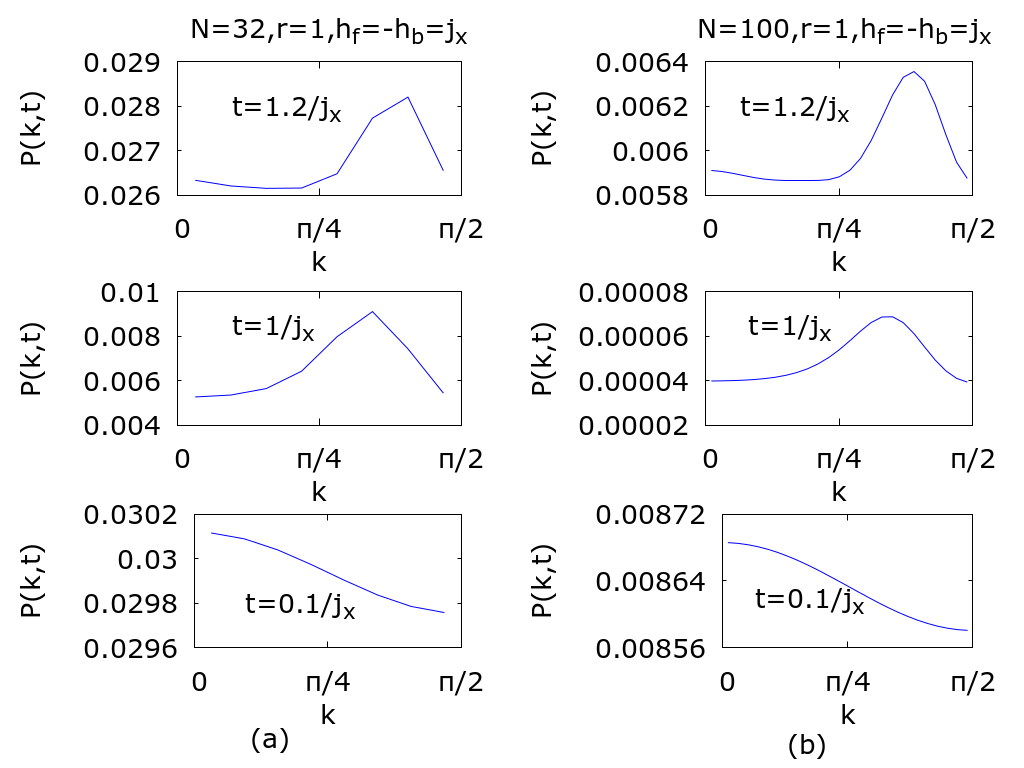}
\caption{The probability distribution function as a function of $k$ at the three different times for the two lengths of the chain. (a) For $N=32$ and (b) For $N=100$ . The initial state is $\ket{\psi(0)}=c_1^{\dag}\ket{00..00}$. The other parameters are set as $h_f=j_x$, $h_b=-j_x$, and $r=1$. The range of $k$ has been shortened to plot just one peak of $P(k,t)$. The position of the peak depends on the time of evolution. We can see the peaks near the same $k$ for $t=1/j_x$ and $t=1.2/j_x$, while for  $j_xt=0.1$ the peak appears flattened towards $k=0$. For a larger spin chain in (b), the characteristic of the momentum distribution remains the same however, the magnitude depends accordingly.}
\end{figure*}

We also show the momentum distribution function for the different time values and also for the different lengths of the spin chain in Fig.10. The distribution function is plotted in the range of $0<k<\pi/2$ for two spin chains of length $N=32$ and $N=100$ in Fig.10(a) and Fig.10(b), respectively. For each chain, we consider three different time values $t=0.1/j_x$, $t=1/j_x$, and $t=1.2/j_x$. The Hamiltonian parameters are $r=1, h_f=j_x$ for forward evolution and $r=1, h_b=-j_x$ for the reverse evolution. In these plots, we have chosen a range of $k$ to show only one peak of the momentum distribution function in any of the plots. Here we can see the time ans momentum dependence of $P(k,t)$. The peaks at $t=1/j_x$, and $t=1.2/j_x$ are centered around the same value of $k=1.9$ for $N=32$ as well as $N=100$, while for $t=0.1/j_x$ the peaks are flattened and have shifted towards the left for both the chains. Also, we can see that the pattern of the momentum distribution remains the same for different $N$. However, the magnitude of $P(k,t)$ has fallen as $1/N$ order. The $P(k,t)$ has magnitude of order $10^{-2}$ at $t=1/j_x$ as compared to the other time values. At this moment the Loschmidt echo is also almost zero as can be seen in Fig.7. 

\section{Loshmidt echo with Kicked Magnetic Field}
\begin{figure}[t]
\includegraphics[height=5cm, width=6cm]{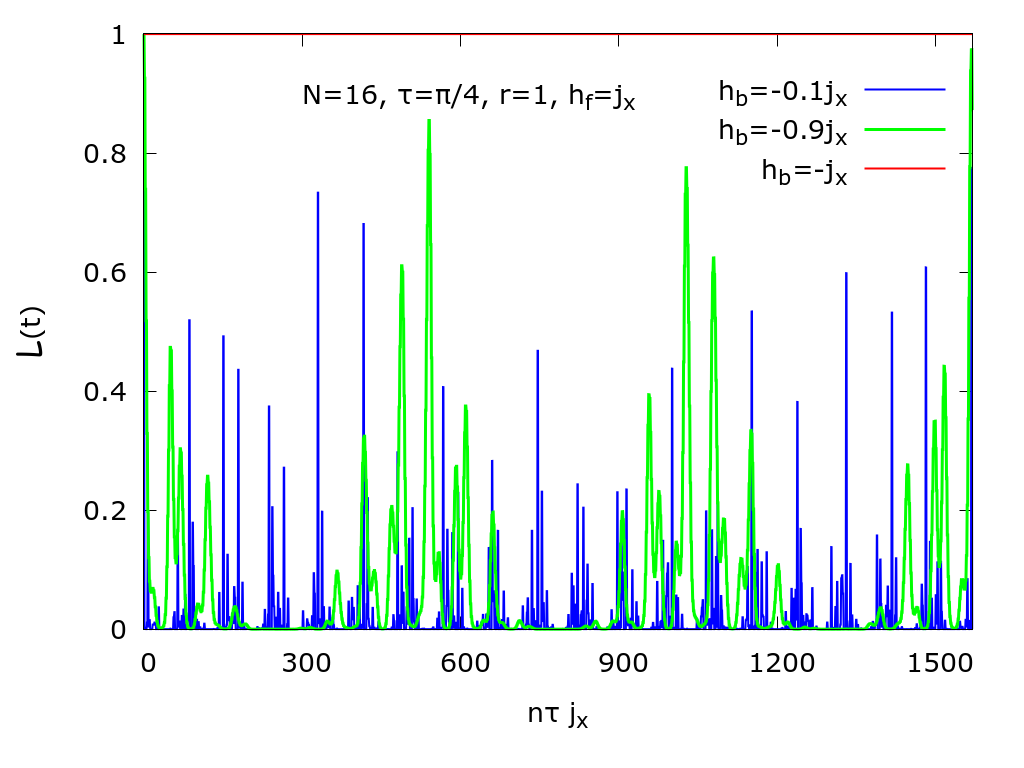}
\caption{The Loschmidt echo for a small chain of $N=16$ at a special kick $\tau=\pi/4$. At $r=1, h_f/j_x=-h_b/j_x=1$, the system does not show the dynamics. Changing the magnetic field to $h_b\rightarrow 0$ in reverse evolution, the LE shows frequent revival peaks. As the magnetic field tuned towards the unity(in $h/j_x\rightarrow 1$) the revival peaks are rare but very significant, which subsequently disappear as $h/j_x=1$.}
\end{figure}
\begin{figure}
\includegraphics[height=5cm, width=6cm]{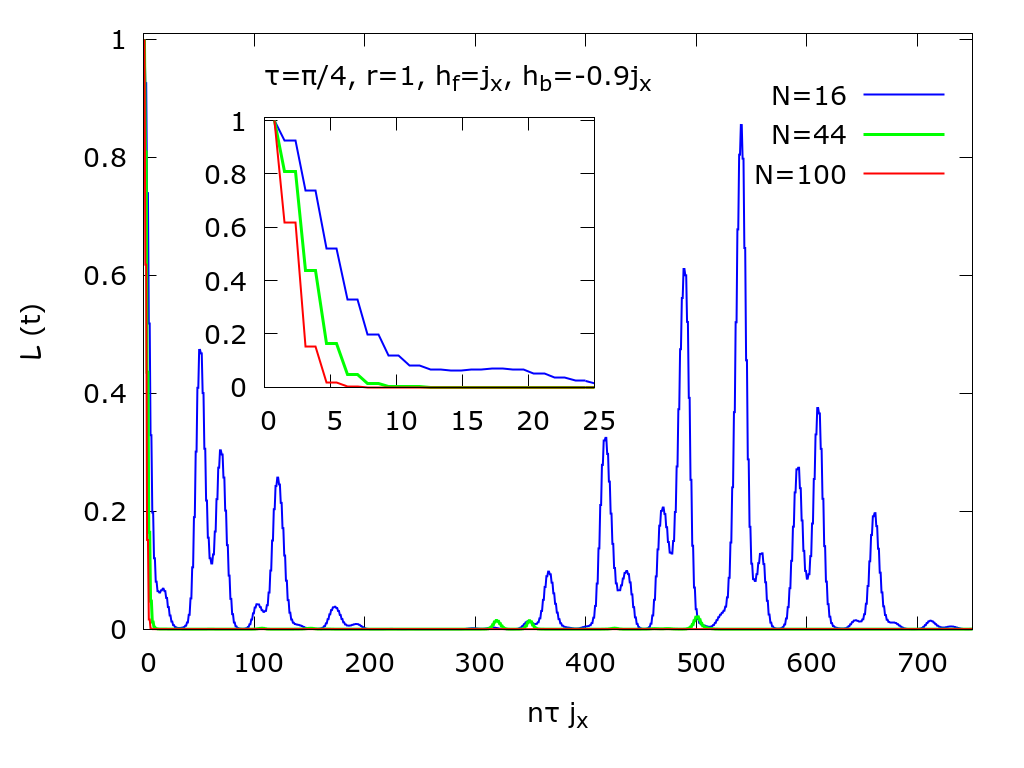}
\caption{Loschmidt echo for different chain lengths at $\tau=\pi/4$. For the larger spin chains, similar to the direct Hamiltonian behavior, the Loschmidt echo falls sharply after the evolution begins, and does not show revival peaks in the long time evolution. However, at $\tau=\pi/4$ and $h_f/j_x=-h_b/j_x=-1$, the LE behavior remains the same.}
\end{figure}
In previous sections, we discussed the Loschmidt echo for the Hamiltonian under the uniform magnetic field. In this section, we consider the interaction Hamiltonian under the kicked magnetic field, which is introduced using the delta function. In general, the characteristic of the dynamics under this Hamiltonian is similar to the dynamics under the constant magnetic field case. However, the kicks values are determining factor in the dynamics. The special values of kicks give surprising results for the Loschmidt echo. The effect of applying the magnetic field using the delta function occurs only at the kicking times, and, in between the two kicks, the dynamics is governed by the critical Hamiltonian, i.e, Hamiltonian with the zero magnetic field. So, the dynamics has a mixed effect of the critical and non critical Hamiltonians. This affects the system most when we choose special kicks values. The analytical solution of the model has been presented in our previous work\cite{Vimal2}. Here, we review for the our convenience. In the symbolic form, the constituent terms of the Hamiltonian in Eq.2 can be written as $H_{xx}= H(j_x,0,0)$, $H_{yy}= H(0,j_y,0)$, and $H_{z}= H(0,0,j_z)$. The full Hamiltonian then is given by $H=H_{xx}+H_{yy}+H_z$. The kicked Hamiltonian can be written as 
\begin{equation}
H=H_{xx}+H_{yy}+\sum_{n=-\infty}^{\infty}\delta(n-\frac{t}{\tau}) h\sum\limits_{i=1}^N\sigma_i^z.
\end{equation}
Here, we apply the magnetic field in the kicked form at period $\tau$. The time after the $n$ kicks are given by $t=n\tau$. The Hamiltonian is periodic over $\tau$. In this case, the dynamics can be governed by the Floquet operator formalism. The Hamiltonian is broken into two parts, namely, the interaction $H_{xx}+H_{yy}$, and the magnetic field $H_z$ Hamiltonians. The unitary operator between the two successive kicks is written as 
\begin{equation}
U=e^{-i\tau (H_{xx}+H_{yy})}e^{-i\tau H_{z}}.
\end{equation} 
And the state after $n$ number of kicks can be written as
\begin{equation}
\ket{\psi_n(t)}=U\ket{\psi_{n-1}(t)}=U^n\ket{\psi(0)}.
\end{equation}
While considering an unentangled state in the computational basis as an initial state, the evolution can only be governed by the interaction Hamiltonian between the two kicks. The unitary operator with the $H_z$ Hamiltonian can give only a phase contribution as the initial state is an eigenbasis of $H_z$. The interaction part of the Hamiltonian can be diagonalized for every mode $q$, which remains associated with the other three momentum values $c_{q-\pi}$, $c_{-q}$, and $c_{\pi-q}$. However, the two parts of the Hamiltonian do not commute, therefore, it is necessary to transform the $H_z$ in terms of the eigenstates of the interaction Hamiltonian(the explicit calculation of this unitary 
\begin{figure}
\includegraphics[height=6cm, width=7cm]{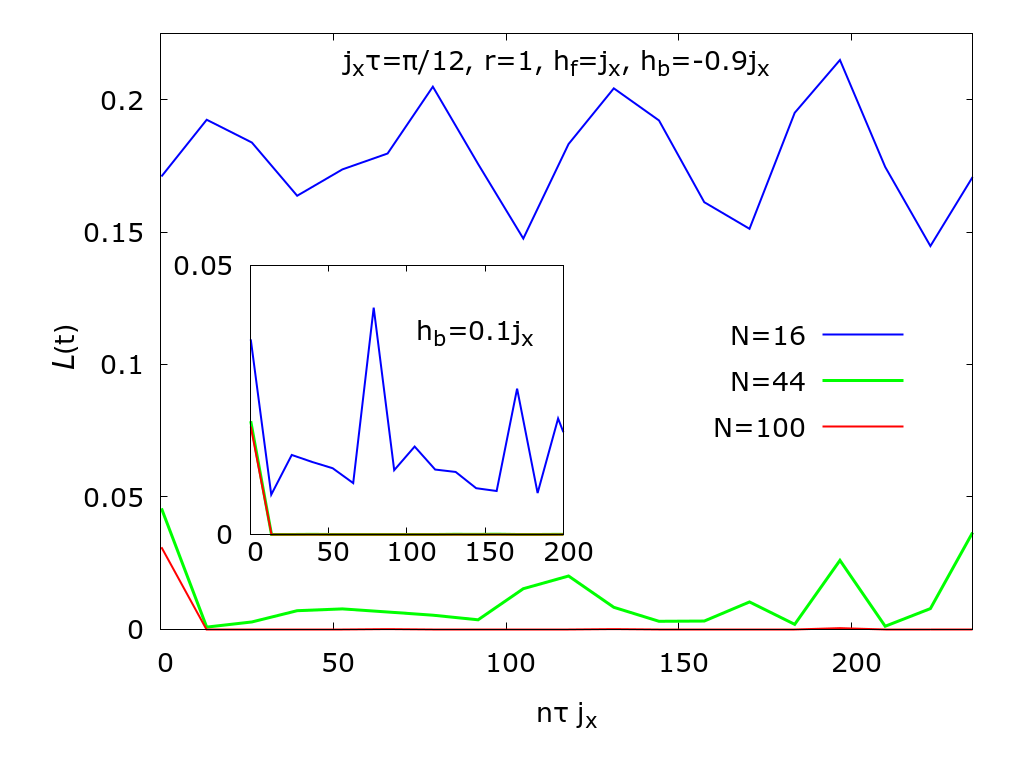}
\caption{Window averaged plots of the Loschmidt echo for the different spin chain lengths. We consider arbitrary $\tau=\pi/12j_x^{-1}$ and the direction of the magnetic field $h/j_x=1$ is reversed in reverse evolution. We can see the LE decreases substantially as length of the spin chain increases from $N=16$ to $N=44$. Also revival peaks are absent for the larger spin chain. In the inset plot, the magnetic field is set near critical point in the reverse evolution, which reduces the Loschmidt echo significantly. Also, the Loschmidt echo saturates at smaller spin system $N=44$, in this case. The window length is considered of 50 kicks.}
\end{figure}
operator can be seen here\cite{Vimal2}). For a mode $q$, the unitary operator in Eq.28 can be written in the tensor product form as $U=V_1\otimes V_2 $. The matrices $V_1$ and $V_2$ can be expressed and diagonalized in the suitable basis states, $\ket{00},\ket{01},\ket{10},\ket{11}$. We recall the eigenvalues of $V_1$, which are written as $\lambda_{\pm}= \frac{1}{2}\left[(e^{4iet}+1)\cos(2ht)\pm\sqrt{{(e^{4iet}+1)}^2{\cos(2ht)}^2-4e^{4iet}} \right] $ with the corresponding eigenstates, $X_{13}^{\dag}\ket{00}= x_1\ket{10}+y_1\ket{01}$, and $Y_{13}^{\dag}\ket{00}=x_2\ket{10}+y_2\ket{01}$, where $x_i,y_i$ are the normalized coefficients of the eigenstates. The subscript labels in $X_{13}\ket{00}=X_{13}\ket{0_10_3}$ are the labels on the fermions for our convenience. Similarly, the eigenvalues of $V_2$ are, $\lambda_{\pm}^{\prime}= \lambda_{\pm}(iet\rightarrow-iet) $, and the corresponding eigenstates are $X_{24}^{\dag}\ket{00}= x_1^{\prime}\ket{10}+y_1^{\prime}\ket{01}$, and $Y_{24}^{\dag}\ket{00}=x_2^{\prime}\ket{10}+y_2^{\prime}\ket{01}$. Thus, the eigenvalues and the corresponding eigenvectors of the unitary $U$ are written as
\begin{equation}
\begin{split}
 \lambda_1=\lambda_+\lambda_+^{\prime},\quad\ket{\lambda_1}=X_{13}^{\dag}X_{24}^{\dag}\ket{vacuum}, \\ 
 \lambda_2=\lambda_-\lambda_+^{\prime},\quad \ket{\lambda_2}=Y_{13}^{\dag}X_{24}^{\dag}\ket{vacuum}, \\ 
 \lambda_3=\lambda_+\lambda_-^{\prime},\quad \ket{\lambda_3}=X_{13}^{\dag}Y_{24}^{\dag}\ket{vacuum},\\
 \lambda_4=\lambda_-\lambda_-^{\prime},\quad \ket{\lambda_4}=Y_{13}^{\dag}Y_{24}^{\dag}\ket{vacuum}.
 \end{split}
\end{equation}
The vacuum state $\ket{vacuum}$ is a direct product of the vacuum states of $V_1$ and $V_2$ matrices. For each eigenvalue $\lambda_i$ in Eq.30, the corresponding eigenvector of $U$ can be rewritten in terms of the momentum operators $c_{q-\pi}$, $c_{-q}$,$c_q$, and $c_{\pi-q}$. We write the first eigenstate of the unitary $U$ as 
\begin{equation}
\begin{split}
\ket{\lambda_1}= &[\alpha_1+\alpha_2(c_q^{\dag}c_{\pi-q}^{\dag})+\alpha_3(c_{-q}^{\dag}c_{q}^{\dag})+ \alpha_4( c_{q-\pi}^{\dag}c_{\pi-q}^{\dag})\\ &+ \alpha_5(c_{q-\pi}^{\dag}c_{q}^{\dag})+\alpha_6(c_{q-\pi}^{\dag}c_{-q}^{\dag}c_q^{\dag}c_{\pi-q}^{\dag})]\ket{0000},
\end{split}
\end{equation}
where the probability amplitudes $\alpha_i=f(x_i,y_i,x_i^{\prime},y_i^{\prime})$ are function of the coefficients of eigenvectors of the matrices $V_1$ and $V_2$. The other eigenstates of unitary operator $U$ can be written in a similar way. It is good to recall that the state $\ket{\lambda_i}$ are written for a mode, and the full state of the system is given by the direct product of $N/4$ such modes as it has been expressed in Eq.7. The eigenstates $\ket{\lambda_i}$ for the kicked Hamiltonian are equivalent to the state $\ket{\phi_q(t)}$ for the direct Hamiltonian case written in Eq.7. For the kicked Hamiltonian, we can write $\ket{\psi_n(t)}$ and $\ket{\psi_n'(t)}$ for the forward and the reverse evolved state, respectively. We follow the same parameters labels like $h_f$ for the magnetic field in forward evolution while $h_b$ for the same in reverse evolution. The state after the $n$ kicks of the magnetic field can be written as

\begin{equation} 
\ket{\psi_n(t)}=U^n\ket{\psi(0)}=\prod_q \sum_{i=1}^{4}\lambda_i^n\ket{\lambda_i}\langle\lambda_i|\psi(0)\rangle.
\end{equation}
Using Eq.32, We compute the Loschmidt echo in the kicked Hamiltonian case for the values of the different parameters. The kicking time of the magnetic field determines the behavior of the Loschmidt echo. For a kick period of infinitesimally small value, the results of the Direct Hamiltonian case and the kicked case are the same. At special kicks, the Loschmidt echo shows surprising behaviors. At $\tau=\pi/4$, with Hamiltonian parameters $j_x=j_y=1$ and $h=j_x$, the wave function does not evolve at all from the initial state $\ket{00..00}$. This can be shown analytically for the spin chain of $N=4$ sites. For other $h/j_x$ values, the wave functions in the forward and the reverse directions show evolution. However, changing the direction of the magnetic field does not alter the wave function, so the overlap of the $\ket{\psi_n'(t)}$ and $\ket{\psi_n(t)}$ gives unity. Also, it is important to see that after the first kick the Loschmidt echo is unity. It is because we have considered the same local interaction parameters $r$ in both directions of evolution. The unitary with the magnetic field only changes the phase of the amplitudes of the wave functions, which in overlap do not change the Loschmidt echo. The Loschmidt echo for a smaller spin chain at $\tau=\pi/4$ is plotted in Fig.11, where it is unity for $r=1$ and $h/j_x=1$. For an arbitrary magnetic field value, it shows revivals at arbitrary times during evolution, and as the magnetic field is tuned towards unity($h/j_x=0.9$) the peaks become more pronounced only at fewer points in the evolution, which all disappears when the magnetic field is tuned at exactly $h=j_x$. In Fig.12, We show the Loschmidt echo for the different spin chains at $\tau=\pi/4$, $j_x=j_y=1$ and $h=j_x$. The reverse field is set at $h_b=0.9j_x$. We can see that in the long-time evolution the smaller chains have revival peaks while the larger spin chains the Loschmidt echo falls exponentially and do not show the revival similar to the Loschmidt echo in the direct Hamiltonian case.

In Fig.13, we show the window averaged  Loschmidt echo at an arbitrary kick $\tau=\pi/12j_x^{-1}$. The window length is set equivalent to 50 kicks. The parameters in the main plot are set as $r=1$, $h_f=j_x$, and the magnetic field is flipped in the reverse direction of evolution. For a smaller chain $N=16$, the revival peaks occur frequently which gives the window average a higher value of nearly 0.2. However, the number of revival peaks of the Loschmidt echo for larger spin chains reduces significantly giving a lower window average. For the spin length as large as $N=100$, the revival peaks of the Loschmidt echo are absent. The inset plot is set at the same parameters as the main plot except that we consider reverse magnetic field $h_b=0.1j_x$. In this case, the Loschmidt echo saturates at the smaller chain length very shortly after the evolution starts, and does not revive even at the relatively smaller length of the chain, $N=44$. 

\section{Conclusion}

We have studied the Loschmidt echo in a Kitaev Hamiltonian under the constant and the kicked magnetic fields. The analytical, as well as the numerical approach, has been applied to compute the Loschmidt echo. We consider different initial states. These states are a no-magnon initial state, a one-magnon initial state with definite momentum, and a one magnon initial state with a uniform probability distribution. The behavior of LE is analytic throughout the evolution at different parameter values. In the direct Hamiltonian case, we have presented results for two parameter sets, one, when the system is near criticality in either direction of the evolution, and second, when the system is away from the criticality in both directions of the evolution. For smaller chains, the Loschmidt echo shows periodically revival peaks in long-time evolution. But unlike quantum correlations like magnetization and concurrences dynamics, the Loschmidt echo does not show the revival peaks in long-time evolution for longer spin chains. Though, it has an short-time revival peak that is present for longer chains. When the system is at the critical point during forward or reverse evolution, the Loschmidt echo does not have such revival peaks. This behavior can also be seen when we show the Loschmidt echo as a function of the magnetic field. However, near the criticality of the system, the Loschmidt echo can have long-time revival peaks even for the larger chains. This behavior is present only when typically small magnetic fields are present in both directions of evolutions. These revival peaks occur at different times contrasting to the short-time revival peaks of the Loschmidt echo in a non-critical regime.

The presence of a magnon with definite momentum in the initial state reduces the revival peaks, which depends on the length of the chain. In longer chains, it does not have a significant effect for a reason that the magnon excitation affects only the evolution of one mode state of the Hamiltonian. Therefore, for sufficiently larger chains, the effect is not seen at all. However, the initial state with equally probable momenta shows a significant drop in subsequent revival peaks even for the longer chains as compared to the revival peaks of the Loschmidt echo with a no-magnon initial state. The presence of momentum in the initial state provides a framework to investigate the probability distributions of different momenta in the evolved state of the system. For the initial state having one-magnon excitation with definite momentum, the probability distribution of a different momentum gives zero even if the momentum belongs to the same mode present in the initial state. The probability distributions for a different momentum in the initial state show the same character during the long-time evolution. However, they have different peak strengths. The probability distribution of the same momentum is equivalent to the Loschmidt echo in this case.

For the initial state of one-magnon with a uniform probability distribution, the momentum distribution function of a mode has four peaks when plotted as a function of momentum. These peaks correspond to the associated momenta of the same mode. 
The momentum distribution is an overlap function with a specific momentum in the initial state, which can give nonzero values when it overlaps with at least four modes states of evolved state in Eq.20, which have that specific momentum excitation. This explains the four-peak structure of the momentum distribution function. At specific times, the probability distribution function as a function of the momentum values attains its maximum near $k\approx 1.2$ for different chain lengths when $N$ is large enough, say near $N=100$. For smaller chain lengths, it differs significantly. The maxima of the probability distribution function fall as the length of the chain increase as $\approx O(1/N)$. This is exactly for the same reason we have for the initial state of one-magnon with definite momentum where the effect of the magnon decreases as the length of the chain increases.

In the last section, we have considered the kicked Hamiltonian case for the Loschmidt echo analysis. The Loschmidt echo generally shows a similar characteristic as the direct Hamiltonian case except at a few special kick parameters. For a larger spin chain, the Loschmidt echo falls to zero in a few kicks just after the evolution and does not revive while for a smaller chain it shows the revival peaks in the long-time evolution. However, in kicked case, the kick parameter may define the character of the Loschmidt echo. One such special kick we consider is $\tau=\pi/4$, where the evolution of the Hamiltonian does not happen at all, for any length of the chain. This behavior is because of the characteristic of the wave function evolving under the kicked Hamiltonian.\\

\section{Acknowledgment} 

VS would like to acknowledge the support of SERB under the matrics scheme.

\end{document}